\documentclass[preprint2]{aastex}

\newcommand{\rs}{r_{\rm s}} \newcommand{\Ms}{M_{\rm s}}
\newcommand{\modotb}{M$_\odot$\ }  
\newcommand{\Mcb}{$M$--$c$\ } 
\newcommand{\rsMsb}{$\rs$--$\Ms$\ } \newcommand{\rsMs}{$\rs$--$\Ms$\ }
\newcommand{\beq}{\begin{equation}} \newcommand{\eeq}{\end{equation}}
\newcommand{\beqa}{\begin{eqnarray}} \newcommand{\eeqa}{\end{eqnarray}} 
\newcommand{\rhoc}{\rho_{\rm s}} 

\usepackage{rotating}    

\slugcomment{Submitted to ApJ, June 2011}

\shorttitle{Halo growth and the NFW profile}
\shortauthors{Vi\~nas et al.}

\begin{document}

\title{Halo growth and the NFW profile}

\author{Jordi Vi\~nas\thanks{E-mail: jvinas@am.ub.es}, Eduard Salvador-Sol\'e and Alberto
Manrique}
\affil{Departament d'Astronomia i Meteorologia, Institut de
Ci\`encies del Cosmos
\\Universitat de Barcelona (UB--IEEC), Mart{\'\i} i Franqu\`es 1,
E-08028 Barcelona, Spain}

\altaffiltext{1}{Associated with the Instituto de
Ciencias del Espacio, Consejo Superior de Investigaciones Cient\'\i
ficas.}

\begin{abstract}
We perform a simple test showing that accreting halos grow inside-out
and have density profiles that are statistically indistinguishable
from those of halos having undergone major mergers. The test also
reveals that the ability for the NFW analytical expression to fit the
density profile of halos spoils towards large masses at any
redshift. On view of these results, we analyze the \Mcb relations on
extreme mass and redshift domains predicted by usual toy models based
on numerical data and one physical model based on the two assumptions
above confirmed by the test.
\end{abstract}


\section{INTRODUCTION}

$N$-body simulations show that dark matter halos have spherically
averaged density profiles that can be fitted by the NFW law
(Navarro et al.~1997)
\beq
\rho(r)=\rhoc\,\frac{4\rs^3}{r\left(r+\rs\right)^2}\,,
\label{NFW}
\eeq
where $\rs$ and $\rhoc$ are the halo scale radius and characteristic
density, respectively. The Einasto profile (Navarro et al.~2004;
Merritt et al.~2005, 2006; Navarro et al.~2010) has also been shown to
give similar or even better fits to halo density profiles. But the
correlations between the NFW shape parameters, namely between
the halo mass $M$ and the concentration $c$, defined as the halo
radius $R$ over $\rs$, or between $\rs$ and the mass $\Ms\equiv
M(\rs)$ encompassed by it, at any given redshift ($z$) still play a
crucial role in the modeling of structure formation.

Much effort has been done in the last decade in accurately determining
these relations from $N$-body simulations (e.g. Zhao et al.~2009;
Klypin et al.~2010; Mu\~noz-Cuartas et al.~2010; Prada et al.~2011 and
references therein). However, such numerical relations are not very
practical because they refer to only a few redshifts and
cosmologies. Moreover, they face the fundamental problem that
numerical simulations have a limited dynamic range, which severely
restricts the mass and redshift domains where they can be inferred. To
circumvent this problem, toy models are often built that fit the
numerical data and are supposed to give acceptable predictions beyond
the domain covered by simulations. 

On the other hand, the NFW law does not exactly fit the halo density
profile, which may introduce poorly known biases in the best-fitting
$\rs$ and $\rhoc$ values depending on the specific halo sample
definition and fitting procedure (including the radial extent)
used. Tasitsiomi et al.~(2004) found indeed that very massive halos at
any given $z$ have density profiles that become closer to a pure
power-law with minus index equal to $\sim 1.5$-2.0. According to these
authors, this would be the consequence that massive halos aggregate
mass at a lower rate, through minor mergers contributing to smooth
accretion, than less massive halos, which suffer major mergers more
frequently. But this interpretation has not been proved. The idea that
the density profile of halos depends on the rate at which they
aggregate mass is often invoked, indeed, to explain some of the
observed trends of halo density profiles (e.g. Wechsler et al.~2002;
Zhao et al. 2003; Mu\~noz-Cuartas et al.~2010). However, other results
of numerical simulations show the opposite result that there is no
particular signature of major mergers in the halo density profiles
(e.g. Huss, Jain \& Steinmetz~1999; Moore et al.~1999; Wechsler et 
al.~2002; Nusser \& Sheth 1999; Romano-D\'iaz et al.~2006).

Clearly, the existence of a reliable physical model of halo structure
could shed light on this apparent conflict and might be used to
predict accurate \Mcb or \rsMs relations at any $M$ and $z$ domain as
well. Salvador-Sol\'e et al.~(2007; hereafter SMGH) built one simple
model that relies on the assumption that {\it halos develop from the
inside out during accretion phases}. The typical density profile for
purely accreting halos is therefore be set by the typical accretion
rate provided by the excursion set formalism, which leads to density
profiles \`a la NFW except at very large masses where they tend to
deviate from the NFW profile in the direction found by Tasitsiomi et
al.~(2004).

The inside-out growth of accreting halos is supported by the results
of $N$-body simulations, which show that halos evolve by conserving
the radial mapping of about 80 \% of their particles (Wang et al.~2010
and references therein). The fact that 20 \% of them do not preserve
their radial mapping might simply be due to the effects of major
mergers yielding the full rearrangement of halos. But this is hard to
prove because of the difficulty to restrict the analysis to purely
accreting halos\footnote{The inside-out growth can also be tested by
directly checking the constancy of the mass of halos inside any given
radius. But this faces the same difficulty: the need to identify
purely accreting objects.}. But even if this were the case, it would
not be clear why the SMGH model is able to predict the right NFW
profile of halos without including major mergers.

Recently, Salvador-Sol\'e et al.~(2011a) have shown from theoretical
arguments that accreting halos must grow, indeed, from the inside out
and that the density profile of purely accreting halos must be
indistinguishable from that of halos having suffered major
mergers. Hence, the typical halo density profile predicted by the SMGH
model is fully justified despite the real growth of these objects both
through accretion and major mergers. Unfortunately, those theoretical
arguments are not straightforward, so the checking against $N$-body
simulations of these important results is necessary.

In the present Letter, we perform one simple test using numerical data
which confirms that: i) halos grow inside-out during accretion phases,
ii) the density profile of purely accreting halos is indistinguishable
from that of halos having suffered major mergers and iii) the fit by
the NFW law of halo density profiles depends on $M$ (at any given
$z$), which causes some trends often misinterpreted as proving that
the profiles of halos depend on whether they suffer major mergers or
not\footnote{This is not to be confused with the fact that the density
profile of individual {\it purely accreting} halos depends, of course,
on their past accretion rate, used in some halo age estimates.}.The
implications of these results on the \Mcb relations predicted far away
from the domain covered by numerical data is then analyzed.

\section{THE TEST}
\label{rsMs}

If halos had density profiles strictly of the NFW form and grew
inside-out by pure accretion, the values of the $\rs$ and $\Ms$
adjusting their density profiles at different epochs would clearly not
vary and the \rsMsb relation itself would remain unchanged. Actually,
halos rearrange in major mergers, so the $\rs$ and $\Ms$ values
suddenly change in those events. But, if the density profiles emerging
from major mergers were indistinguishable from those of purely
accreting halos, the whole \rsMsb relation should be kept unaltered
regardless of the aggregation history of halos. That is, it should be
time-invariant. Moreover, as the points of the \rsMsb relation would
not depend on halo mass, the \rsMsb curve should not privilege any
particular mass. Hence, the \rsMsb relation should also be scale-free.

Conversely, if accreting halos did not have density profiles of the
NFW form or they did not grow inside-out or still the profile for
purely accreting halos were distinct from that of halos suffering
major mergers, the $\rs$ and $\Ms$ values would change as they evolve
and the \rsMs relation would vary with $z$. Moreover, as major mergers
dominate the growth of low-mass halos and smooth accretion that of
very massive ones, the \rsMs relation would not be scale-free.

Thus, the time-invariant, scale-free behavior of the \rsMsb relation
is an unambiguous proof of the inside-out growth of accreting halos
and the similarity of density profiles for purely accreting and
merging halos. Hence, it is an unambiguous proof of the validity of
the SMGH model (as far as the halo mass growth is reasonably
well-described by the excursion set formalism).

\begin{figure}[t]
 \includegraphics[scale=0.38]{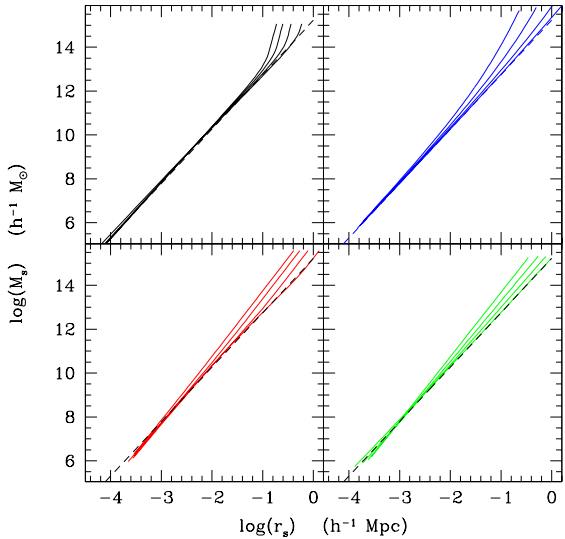}
 \caption{Typical \rsMsb relations at $z=0$, 1, 2 and 3 according to
 Zhao et al.~(2009) (red lines), Klypin et al.~(2010) (blue lines) and
 Mu\~noz-Cuadras et al.~(2010) (green lines) and predicted by the SMGH
 model (black lines). The straight dashed black lines trace the ideal
 \rsMsb relation defined in section \ref{Mcb}.}
\label{msrs2}
\end{figure}

In Figure \ref{msrs2}, we show the \rsMsb relations drawn from the toy
models of \Mcb relations (see Sec.~\ref{Mcb}) proposed by Zhao et
al.~(2009; top-right panel), Klypin et al.~(2010; bottom-left panel)
and Mu\~noz-Cuartas et al.~(2010; bottom right-panel) fitting
numerical data in their simulations of the same concordance cosmology
with $(\Omega_m,\Omega_\Lambda,h,\sigma_8)=
(0.27,0.73,0.70,0.82)$. (Mu\~noz-Cuartas et al. actually use a
slightly different concordance model, but this makes no difference at
the resolution of the Figure.) The curves at $z=0$ are very close to
straight lines in log-log as expected, but, for increasing $z$, they
progressively bend upwards, so they do not overlap.

Does this mean that the SMGH model is wrong? Not really. In the same
figure (top-left panel), we plot the \rsMs curves predicted from the
SMGH model. Again, the log-log relation at $z=0$ is very approximately
a straight line, but, as $z$ increases, the relations progressively
bend upwards and the curves do not overlap. As in the SMGH model all
halos accrete (there is no major merger) and develop, by construction,
from the inside out, this can only be due to the fact that they do not
strictly have density profiles of the NFW form. Indeed, even if the
density profile of a halo does not vary during accretion, its fit by
the NFW law will then result in best values of $\rs$ and $\Ms$ that
slightly shift as the radial extent of the halo (setting the
fitting radial range) increases.

\begin{figure}[t]
 \vspace{-2.5 cm}
\centerline{
 \includegraphics[scale=0.38]{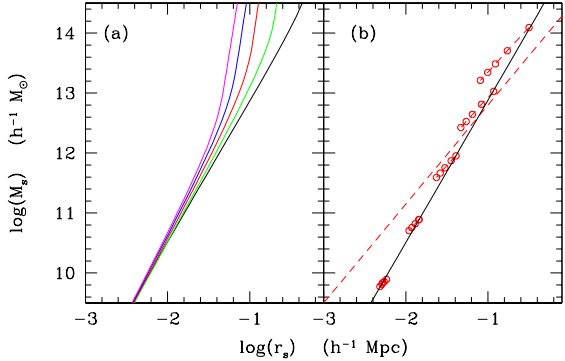}
}
 \caption{Panel (a): \rsMsb relation (full lines) predicted by the
 SMGH model at $z=0$, 2, 4, 6 and 8 (respectively in black, green,
 red, blue and magenta).\ Panel (b): Tracks followed by typical
 accreting halos (dashed red lines) with current masses equal to
 $10^{11}$, $10^{12}$, $10^{13}$, $10^{14}$ and $10^{15}$ \modotb
 (from bottom to top) and the same redshifts (circles) as in the left
 panel. In red dashed straight line, the universal direction followed
 by slowly accreting halos found by Zhao et al.~(2009). In full black
 line, the a straight line with the same slope as the \rsMsb relation
 predicted by the SMGH model at $z=0$.}
\label{msrs1}
\end{figure}

This interpretation of the actual behavior of the \rsMs relation is
confirmed by the results plotted in Figure \ref{msrs1}. In panel (a),
we show again the \rsMs relations predicted at several $z$'s by the
SMGH model but in a somewhat smaller $\rs$ range and in panel b) we
show the tracks followed for decreasing $z$ by a few individual halos
tracing such \rsMs relations. In the SMGH model, all halos develop
inside-out. Yet, the halos still move in the \rsMsb plane along one
universal direction due, as checked, to the effect just
mentioned. Remarkably, the slope of that universal shift is identical
to that ($y=x^{3\times0.55}$) of the shift also undergone by halos in
the simulations by Zhao et al.~(2009) (see their Figure 22). Moreover,
the shifts found by Zhao et al. end up at $z=0$ along the same
straight line tracing the SMGH \rsMsb relation at the same $z$. This
unambiguously proves that such shifts have the same origin in the SMGH
model as in Zhao et al. simulations. Zhao et al. interpreted them as
due to a progressive change in the inner halo density profile during
slow accretion phases (see also Zhao et al.~2003). However, as
revealed by the SMGH model, the inner density profile for these halos
does not actually change; the small shift in the \rsMsb plane is due
to the varying best-fitting $\rs$ and $\Ms$ values obtained as the
halo radial extent increases due to the fact that at massive halos are
not well-fitted by the NFW profile.

When this effect is corrected for, that is, when the values of the
shape parameters adjusting the density profile at any $z$ are replaced
by those adjusting the same profile extended so to encompass the whole
radial range at $z=0$, then the log-log \rsMsb relations at any $z$
become straight lines that overlap. In other words, the test confirms
that accreting halos evolve inside-out and that the density profiles
of purely accreting halos are statistically indistinguishable from
those of halos having suffered major mergers.

\section{IMPLICATIONS FOR THE \Mcb RELATION}
\label{Mcb}
 
Integrating $4\pi r^2$ times the NFW profile (eq.~[\ref{NFW}]) written
in terms of $\Ms$ instead of $\rhoc$ and taking into account the
relation $\Ms(\rs)=A\rs^\nu$ with $A=9.99\times 10^{14}$ \modotb
Mpc$^{-\nu}$ and $\nu=2.481$ fitting the numerical \rsMsb relations at
$z=0$, one is led to
\beqa
M^{3-\nu}(c,z)=\left[\frac{4\pi}{3} \Delta_{\rm vir}(z)\bar\rho(z)\right]^{-\nu} ~~~~~~~~~~~~~~~~\nonumber\\\times\left\{\frac{A c^{-\nu}}{\ln(2)-0.5}\left[\ln(1+c)-\frac{c}{1+c}\right]\right\}^3\,,
\label{Mcz}
\eeqa
where $\Delta_{\rm vir}(z)$ and $\bar\rho(z)$ are respectively the
halo overdensity and mean cosmic density at $z$. Equation (\ref{Mcz})
therefore defines the \Mcb relation {\it for halos strictly endowed
with the NFW profile (in particular at $z=0$), growing inside-out
during accretion periods and with no signature of major mergers}. Such
an \Mcb relation is from now on called ``the ideal \Mcb relation''.

\begin{figure}[t]
 \includegraphics[scale=0.38]{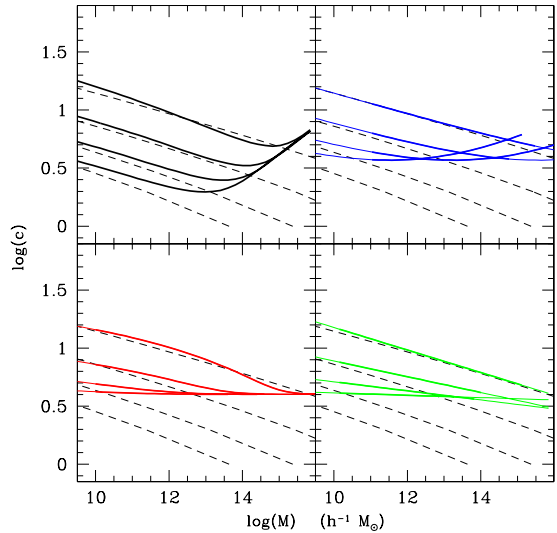}
 \caption{Comparison between the \Mcb relations predicted by the SMGH
 model (full black lines) and proposed by Zhao et al.~(2009) (red
 lines), Klypin et al.~(2010) (blue lines) and Mu\~noz-Cuadras et
 al.~(2010) (green lines) at $z=0$, 1, 2 and 3 in the same cosmology
 as in Figure \ref{msrs2}. We distinguish that part of the curves
 covered by simulations (thick colored lines) and their
 extrapolations outside it (thin colored lines). The ideal NFW \Mcb
 relation (see text) is in dashed black lines.}
\label{mc1}
\end{figure}

In Figure \ref{mc1}, we compare such an ideal \Mcb relation with the
\Mcb relations predicted by the SMGH model and the toy models by Zhao
et al.~(2009), Klypin et al.~(2010) and Mu\~noz-Cuartas et al.~(2010)
derived from numerical data, respectively given by
\beq
 c(M,z)=\left\{4^8+\left[\frac{t(z)}{t_{0.04}(M)}\right]^{8.4}\right\}^{1/8}\,,
\label{Zhao}
\eeq
with $t(z)$ equal to the cosmic time corresponding to $z$ and
$t_{0.04}(M)$ the time when the main halo progenitor gained $4\%$ of
its mass $M$ at $z$,
\beqa
c(M,z)=9.2\,\delta^{1.3}(z)\left(\frac{M}{10^{12}h^{-1}M_{\odot}}\right)^{-0.09}
~~~~~~~~~~\nonumber\\\times\left\{1+0.013\left[\frac{M}{10^{12}h^{-1}M_{\odot}}\delta(z)^{-\frac{1.3}{0.09}}\right]^{0.25}\right\}\,,
\label{Klypin}
\eeqa
and
\beq
c(M,z)=10^{b(z)}\,\left(\frac{M}{h^{-1} M_\odot}\right)^{a(z)}
\label{Maccio}
\eeq
where $\delta(z)$ is the critical overdensity for collapse at $z$,
$a(z)\equiv 0.029z-0.097$ and $b(z)\equiv
-110.001/(z+16.885)+2469.72/(z+16.885)^2$.

All the curves drawn from numerical data behave in a similar way.
Contrarily to the \rsMs relation, they show a strong dependence on
$z$. At $z=0$ they closely follow the ideal \Mcb relation but, at
larger $z$, they progressively deviate from it at large masses likely
due to the effect mentioned in section \ref{rsMs}. In fact, for large
enough $M$, the ideal decreasing trend brakes and the empirical curves
level off or even begin to increase again (in the case of Klypin et
al.~2010; see also Prada et al.~2011). The \Mcb relation predicted by
the SMGH model shows similar trends although it remains closer to the
ideal \Mcb relation until slightly larger $M$ at each $z$. This might
reflect a slight flaw of the SMGH model at very large $M$. But the
possibility cannot be ruled out that the empirical \Mcb relations are
somewhat biased owing e.g. to the impact of using non-fully relaxed
halos (more abundant at large masses) or of estimating the typical
halo concentration in mass bins by means of the median
value.

\begin{figure}[t]
 \includegraphics[scale=0.38]{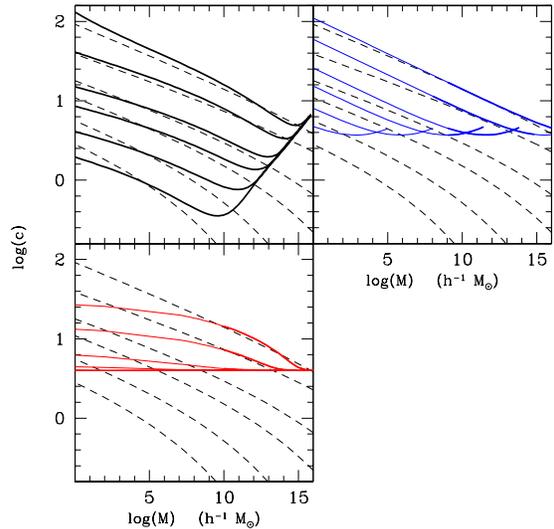}
 \caption{Same as Figure \ref{mc1} (and identical symbols) but for a much
 wider mass and redshift range. The different curves in each panel
 are for $z$ equal to 0, 1, 3, 5, 9, 15 (from top to bottom on the left
 of each panel).}
\label{mc2}
\end{figure}

In Figure \ref{mc2}, we plot the same \Mcb relations extended over a
much wider domain ($M$ down to 1 \modotb and $z$ up to 15). We exclude
the \Mcb relations by Mu\~noz-Cuartas et al.~(2010) because these
authors state explicitly that their toy model is a fitting expression
that cannot be extrapolated. As can be seen, the two sets of
extrapolated empirical \Mcb relations have a lower bound at about
$log(c)\sim 0.6$ contrarily to the ideal \Mcb relations. However, in
Klypin et al.~(2010), the curves increase again after reaching that
minimum, whereas, in Zhao et al.~(2009), they level off. Notice that
this discrepancy affects the part of the curves that adjusts numerical
data, meaning that the disagreement between both groups is very
relevant. At small $M$, the two sets of curves diverge from each other
even more markedly: in Zhao et al. they show flat asymptotes, while in
Klypin et al. they have increasing asymptotes similar to the ideal
\Mcb relations, although their dependence with $z$ is substantially
less marked.

The $z$-dependence of the SMGH curves is instead quite similar (for
intermediate $M$ at each $z$) to that of the ideal \Mcb
curves. Moreover, the $M$-dependence at any $z$ is also much closer to
that of the ideal \Mcb relations than for the extrapolated empirical
relations. They only substantially deviate from each other at very
small $M$. In brief, the SMGH \Mcb relations are globally in better
agreement with the ideal \Mcb relations than the pseudo-empirical \Mcb
relations. 

Certainly, the ideal relations must not necessarily trace the \Mcb
relations of real halos because, as shown, their density profiles are
not exactly fitted by the NFW law. This translates into an apparent
variation of $\rs$ and $c$ with increasing $z$, which is not present
in the ideal relations. Thus, on the sole basis of the previous
comparison, it is hard to conclude which \Mcb relation is the most
reliable for real simulated halos in such extreme $M$ and $z$
domains. As mentioned, the SMGH model might be affected by some slight
flaw at the large mass end (and perhaps at the small mass end as well)
at any given $z$. But the empirical curves might be biased too, as
suggested by their distinct behavior at very large masses where
numerical data are available. More importantly, their extrapolation at
very small $M$ show so different behaviors at any $z$ that they cannot
be trusted. Last but not least, Salvador-Sol\'e et al.~(2011b) have
recently shown that the properties of halo substructure found in
high-dynamic range $N$-body simulations (Springel et al.~2008), very
sensitive to the behavior of the \Mcb relation in an $M$ and $z$
domain as wide as the one considered here, clearly favor the SMGH
model in front of the toy models by Klypin at al. and Zhao et al.

\section{SUMMARY}

We have performed a simple test showing unambiguously that accreting
halos grow inside-out and that the density profiles of halos grown by
pure accretion are indistinguishable from those of halos having
suffered major mergers. The variation in the NFW shape parameters as
halos accrete is due to the fact that the density profile for very
massive halos is not well-fitted by the NFW law, which causes the
best-fitting values of $\rs$ and $\rhoc$ (or $\Ms$) to slightly vary
as the radial extend of halos increases. After correction of this
effect, the \rsMs relation is linear in log-log and overlaps for all
$z$ as expected.

Over very wide $M$ and $z$ domains, the ideal \Mcb relations for halos
with exact NFW profiles and evolving inside-out by accretion deviate
notably both from the extrapolated \Mcb relations inferred from
numerical data as well as from the \Mcb relation predicted by the SMGH
model. This difference is once again partly due to the fact that the
density profile for simulated and modeled halos is strictly not of the
NFW form. However, the \Mcb curves predicted by the SMGH model stay
closer to such ideal \Mcb curves over a wider $M$ and $z$ domain than
the extrapolated \Mcb relations drawn from toy models, which also
notably deviate from each other. This suggests the better behavior of
the SMGH \Mcb relations at such extreme $M$ and $z$ domains, a
conclusion that meets that found on the basis of completely
independent arguments by Salvador-Sol\'e et al.~(2011b).

\vspace{0.75cm} \par\noindent
{\bf Acknowledgments} \par

\noindent This work was supported by the Spanish DGES
AYA2009-12792-C03-01. We thank Steen H. Hansen for useful comments and
discussions. One of us, JV, beneficiary of the grant BES-2007-14736,
thanks the Dark Cosmology Centre for facilities during his stage at
this center.

\end{document}